\documentclass[runningheads]{svmult}

\usepackage{makeidx}   
\usepackage{graphicx}  
\usepackage{subeqnar}  
\usepackage{multicol}  
\usepackage{physprbb}  
\makeindex             

\newcommand{\greeksym}[1]{{\usefont{U}{psy}{m}{n}#1}}
\newcommand{\umu}{\mbox{\greeksym{m}}}

\begin{document}
\title*{The SWIRE SIRTF Legacy Program:\protect\newline 
Studying the Evolutionary Mass Function and Clustering of Galaxies}
\toctitle{The SWIRE SIRTF Legacy Program:
\protect\newline studying the evolutionary mass function and clustering of galaxies}
%
%
\titlerunning{The SWIRE SIRTF Legacy Program}
%
\author{Alberto Franceschini\inst{1},
Carol Lonsdale\inst{2}, 
\and the SWIRE Co-Investigator Team
}
\authorrunning{A. Franceschini \& C. Lonsdale}
%
%
\institute{Astronomy Department, Padova University, I-35122
Vicolo Osservatorio 2, Padova, Italy
\and 
IPAC, Pasadena, USA}

\maketitle              

\begin{abstract}

The SIRTF Wide-area Infrared Extragalactic (SWIRE) survey is a {\sl Legacy Program}
using 851 hours of SIRTF observing time to conduct a set of large-area ($\sim 67$ 
square degrees split into 7 fields) high Galactic latitude imaging surveys, 
achieving 5-sigma sensitivities of 0.45/2.75/17.5 mJy at 24/70/160 
$\mu$m with MIPS and of 7.3/9.7/27.5/32.5 $\mu$Jy at 3.6/4.5/5.8/8.0 $\mu$m with IRAC.
These data will yield highly uniform source catalogs and high-resolution calibrated 
images, providing an unprecedented view of the universe on co-moving scales up to 
several hundreds Mpc and to substantial cosmological depths ($z\simeq 2.5$
for luminous sources). SWIRE will, for the first time, study evolved stellar systems 
(from IRAC data) versus active star-forming systems and AGNs (from MIPS data) in the 
same volume, generating catalogues with of order of 2 million infrared-selected galaxies. 
These fields will have extensive data at other wavebands, particularly in the optical, 
near-IR and X-rays. 
SWIRE will provide a complement to smaller and deeper observations in the SIRTF 
Guaranteed Time and the Legacy Program GOODS, by allowing the investigation of the
effect of environment on galaxy evolution.
We expand here on capabilities of SWIRE to study with IRAC the evolution 
of the bright end of the galaxy mass function as a function of cosmic time.

\end{abstract}

\section{Introduction}
Measuring the mass function of distant galaxies and its evolution with cosmic time is
an ultimate task when attempting to recover the origin of the presently observed cosmic 
structure. The study of this integral of the stellar mass content as a function of time 
makes a complementary approach to the usual determination of the instantaneous rate of
star formation, and has the advantage over the latter of not being sensitive at all to 
the effects of dust extinction, otherwise a quite fundamental limitation.
In particular, the evolution of the massive end of the galaxy population, tracing the 
assembly of present-day luminous galaxies, is a critical 
cosmogonic observable.

Current large optical/NIR telescopes allow in principle the measurement of
dynamical masses through high spatial and spectral resolution spectroscopy of 
selected samples of high-redshift galaxies (e.g. Koo;
Rigopoulou et al., these Proceedings). However, the long time integrations
needed, and the difficulty to sample all but the inner galaxy cores due to the
cosmological surface brightness dimming, hamper systematic studies of sizeable galaxy 
samples over large areas.

\begin{figure}[]
\begin{center}
\includegraphics[width=0.75\textwidth,height=5.4 cm]{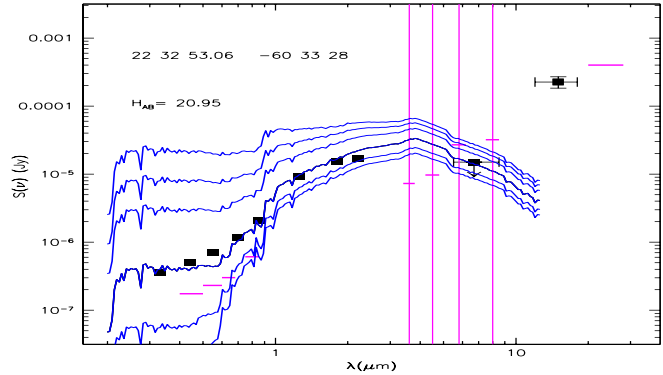}
\end{center}
\caption[]{Synthetic spectra of galaxies as a function of the age of the
stellar population. Ages are 1, 2, 3, 5, 7 and 9 Gyr from top spectrum to bottom. 
The synthetic spectra were computed with the PEGASE code (Fioc \& Rocca-Volmerange
1997), and assume a Salpeter IMF, the Padova stellar tracks, an 
exponentially decreasing star-formation rate
with 1 Gyr time-scale, no extinction and no nebular emission.
The 5 Gyr spectrum is fit to photometric data on a z=1.27 ISO-detected galaxy
in the HDF South. 
The vertical lines with horizontal ticks indicate the $\lambda_{eff}$
and 5$\sigma$ sensitivities of SWIRE.
%
}
\label{eps1}
\end{figure}

Substantial sky areas and numerous samples would be required on consideration 
that galaxy evolution is evidently a function of the environment.
Although the details may still be missing, there are indications
that the amount of evolution of massive galaxies in rich clusters 
is different from what could have happened in the galaxy field (e.g. Franceschini et al. 
1998; Stanford et al. 1998). In general, old red galaxies appear to cluster more
strongly than the younger blue galaxy population.
So any conclusions based on even deep investigations of small sky areas have to
be considered as tentative untill they are proven in representative cosmic volumes.

In principle, a powerful alternative to the time-expensive spectroscopic investigations for the
study of the evolutionary mass function exploits observations of the spectral intensity 
of galaxies in the near-IR and its weak dependence on the age of the contributing stars
(compared with that of the optical SED).
An illustration of the robustness of such determination of the baryonic mass for an 
ISO-discovered galaxy at z=1.27 is given in Fig. \ref{eps1}. 
The same figure also shows, however,
that sampling the emission longwards of the athmospheric limit at $\lambda=2.5\ \mu$m
is needed for a proper mass determination, particularly at such high or higher
redshifts.

SIRTF, to be launched end of 2002, will soon allow imaging with unprecedented 
sensitivity in this poorly known part of the e.m. spectrum. 
This will be achieved with the Infrared Array Camera (IRAC), one of the
three SIRTF science instruments, providing simultaneous 
5.12x5.12 arcmin images at 3.6, 4.5, 5.8, and 8 $\mu$m, using two InSb 
and two SiAs 256x256 detector arrays.

\section{The SWIRE Project}

The SWIRE survey is the largest SIRTF Legacy Program, devoted
to image the evolution of dusty star-forming galaxies, evolved stellar
populations, and AGNs as a function of environment, from redshifts
$z\sim 2.5$ to the current epoch (Lonsdale et al. 2001).  Building on ISO' heritage, 
SWIRE complements smaller and deeper Guaranteed Time SIRTF surveys, and
paves the way for Herschel-FIRST (50$<\lambda<670 \mu$m) and later NGST 
(0.5$<\lambda<30 \mu$m).

The key scientific goals of SWIRE are to view infrared emission from cosmic structures
in the mid- and far-infrared, 
within the key redshift range $0.5<z<2.5$ where much of their formation has occurred.
The main themes under study will be:

a) The evolution of active star-forming galaxies and passively evolving systems 
in volumes large enough to place it in the context of structure 
formation and to study the effect of galaxy environment.

b) The spatial distribution and clustering of starburst galaxies, evolved galaxies, 
and AGN, and the clustering evolution.

c) The identification and characterization of rare classes of sources, like the most 
luminous powerhouses at high-redshifts (the Ultra-luminous
and Hyper-luminous IR sources) and the reddest (presumably the oldest, most
massive and highest redshift) galaxies.

d) The evolutionary relationships between galaxies and AGN, and the contribution of 
accretion energy from obscured AGNs to the cosmic backgrounds.

\vskip -5.3pt 
\begin{table}
\caption{SWIRE Sensitivities and Confusion Limits}
\begin{center}
\renewcommand{\arraystretch}{1.4}
\setlength\tabcolsep{5pt}
\begin{tabular}{llllllll}
\hline\noalign{\smallskip} 
\hline
Band ($\mu$m)               & 3.6    & 4.5    & 5.6    & 8      & 24   & 70   & 160  \\
Sensitivity [mJy], 5$\sigma$& 0.0073 & 0.0097 & 0.0275 & 0.0325 & 0.45 & 2.75 & 17.5 \\
Confusion$^{\mathrm a}$ [mJy], 5$\sigma$  
                            &$4\ 10^{-5}$&$1.5\ 10^{-4}$&$4.3\ 10^{-4}$& 0.001  & 0.085& 3.7  & 36.  \\
\hline\noalign{\smallskip} 
\hline
\end{tabular}
\end{center}
$^{\mathrm a}$ Based on models by Franceschini et al. (2001), and paper in preparation.
\label{Tab1}
\end{table}
\vskip -10.3pt

\subsection{SWIRE Survey Design: Depth and Area Coverage}

Various important cosmological topics have remained untouched even after the ISO 
mission and the deep surveys with large millimetric telescopes.
These include probing the IR emissivity properties of cosmic sources at $z>1$,
detection at $\lambda>3\mu$m of the redshifted stellar continuum by high-z galaxies,
the clustering properties of IR sources, the effects induced by the environment,
and the characterization of populations of rare sources at high-z.
While the GOODS survey (M. Dickinson, these Proceedings) will be very effective in 
addressing the former two items,
the latter will require the large survey area capability provided by SWIRE.

\begin{table}
\caption{SWIRE Survey Areas}
\begin{center}
\renewcommand{\arraystretch}{1.3}
\setlength\tabcolsep{3pt}
\begin{tabular}{llllll}
\hline\noalign{\smallskip} 
\hline
Field name & Field centers &  $\beta$ & I($100\mu m$) & E(B-V) & Area       \\
           &  (J2000)      &  [deg.]& [MJy/sr]    &        & [sq.deg.] \\
\noalign{\smallskip}
\hline
\noalign{\smallskip}
ELAIS-S1      & 00h38m30s -44d00m00s &$-43$ & $0.42$ & 0.008 & 14.8 \\
XMM-LSS       & 02h21m00s -05d00m00s &$-18$ & $1.30$ & 0.027 & 9.3 \\
CHANDRA-S     & 03h32m00s -28d16m00s &$-48$ & $0.46$ & 0.001 & 7.2  \\
Lockman Hole  & 10h45m00s +58d00m00s &$+44$ & $0.38$ & 0.006 & 14.8 \\
Lonsdale Field& 14h41m00s +59d25m00s &$+68$ & $0.47$ & 0.012 & 6.9 \\
ELAIS-N1      & 16h11m00s +55d00m00s &$+74$ & $0.44$ & 0.007 & 9.3 \\
ELASI-N2      & 16h36m48s +41d01m45s &$+62$ & $0.42$ & 0.007 & 4.5  \\
\hline\noalign{\smallskip} 
\hline  
\end{tabular}
\end{center}
\label{Tab2}
\end{table}

The depth of the SWIRE survey (see Table \ref{Tab1}) was thus designed to probe 
over the largest possible sky area the $\sim0.5<z<2.5$ redshift interval, where 
much evolution must have occurred,  
so that the highly transient ($\sim10^{7-8}$ yr) starburst 
and AGN phenomena can be studied in the
context of the major galaxy populations at the same epoch. 
The area of each SWIRE field was defined to provide coverage of angular scales
in the sky corresponding to linear scales of $\sim 50$ to 100 Mpc at $z>1$.
Seven such separate target fields will be surveyed to take care of the noise from 
cosmic variance.

The SWIRE survey fields (see Table \ref{Tab2}) 
were chosen using the  COBE-normalized IRAS 100 $\mu$m 
maps by Schlegel et al. (1998) as areas in the sky with: $I(100)<0.4$ MJy/sr; 
ecliptic latitude $|b|>40^{\circ}$; and contiguous
area $>8^{\circ}$. Within these, additional constraints were imposed by
minimum cirrus contamination, absence of bright stars and galaxies, absence
of bright radio sources and nearby galaxy clusters, and guided by availability 
of multi-wavelength supporting data. 
SWIRE also includes 10 contiguous sq.deg. in the XMM-LSS survey area that is 
going to be covered with XMM in the GTO and GO program.

\subsection{Multi-band Optical/NIR Identification Program}

Complementary data at other wavelengths are needed for source identification and
physical characterization. The optical/near-IR follow-up consists of various
programs. The entire 70 sq.deg. survey is being imaged to $r' \sim 25$, to 
identify $\sim 75\%$ of the SWIRE population.
Multi-band imaging ($g'\sim 26$, $r'\sim 25$, $\sim i'~24$, $K_s\sim 19.5$)
over partial fields ($\sim 4$ sq.deg.) in several of the 7 survey fields will
provide photometric redshifts. 
Deeper imaging ($g'\sim 27$, $r'\sim 26$, $\sim i'~25$) over 1 sq. deg. in a couple 
of fields, and in particular the ESO Large Program ESIS for ELAIS-S1 
(B=26, V,R=25.5, I=25.5, Z=24 over 6 sq.deg., 
see http://www.eso.org/observing/visas/lpsummary/168.A-0322.html)
will allow the characterization of the
faintest and highest-z SWIRE source populations, and the identification of 
high-redshift clusters and groups.

Radio data at moderate depths from ATCA are available on 4 sq.deg. of ELAIS-S1,
(see Gruppioni et al. 1999), while a very deep VLA survey has
been approved for the Lockman Hole area.
Shallow X-ray XMM data may eventually cover the whole XMM-LSS field, 
while deeper data may become available for the inner LSS part,
and for other fields, should pending proposals be accepted.

\begin{figure}[]
\begin{center}
\includegraphics[width=.7\textwidth,height=5.5 cm]{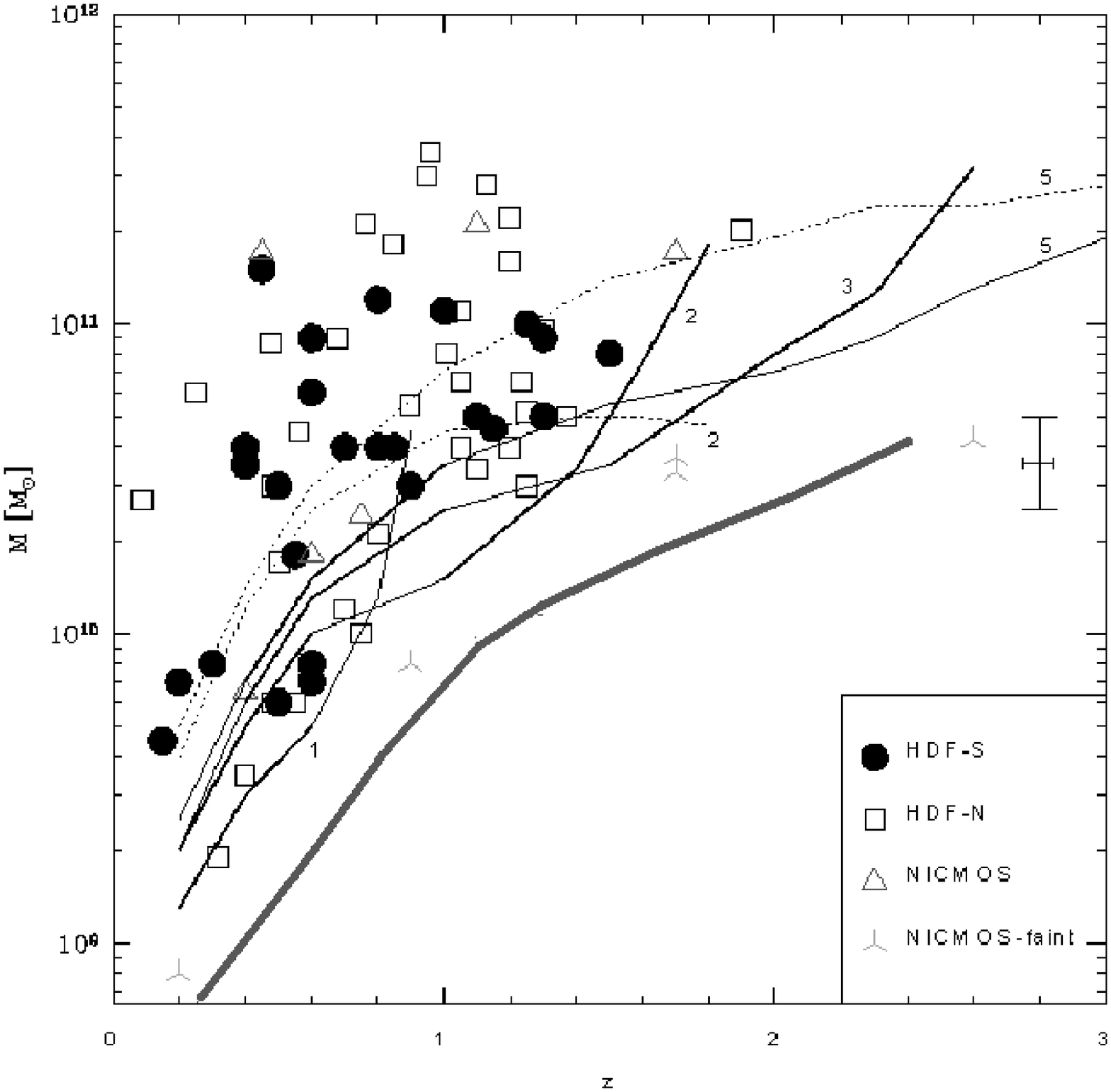}
\end{center}
\caption[]{Masses of morphologically-selected E/S0 galaxies with K$<20.15$, based on 
fits to the optical/NIR SEDs, vs. redshifts (Rodighiero et al. 2001). 
The thin lines correspond to the K-band flux limit for different 
photometric models. The lower thick line corresponds to the equivalent "mass limit" 
for the SWIRE/IRAC surveys.
}
\label{eps2}
\end{figure}

\subsection{SWIRE Data Products and Archiving Services}

The SWIRE data products that will be released to the community on a 6-month 
timescale from observation will consist of preliminary -- though highly 
reliable and complete -- source lists, FITS images, cross-band 
identifications, sky coverage maps and documentation, as well as 
ancillary data that are available to a similar level of validation 
by the delivery date. Successive deliveries will take care of improving the
area coverage, sensitivity limits, mosaicing, and cross-matching over more and
more observing bands (including ancillary data).

The SWIRE dataset will be served through the IPAC InfraRed Science Archive (IRSA), 
which supports extensive data-analysis services, including fast catalogue and image 
sub-setting and correlation (see http://ipac.caltech.edu/IRSA/ for a preview).

\begin{figure}[]
\begin{center}
\includegraphics[width=.8\textwidth,height=5. cm]{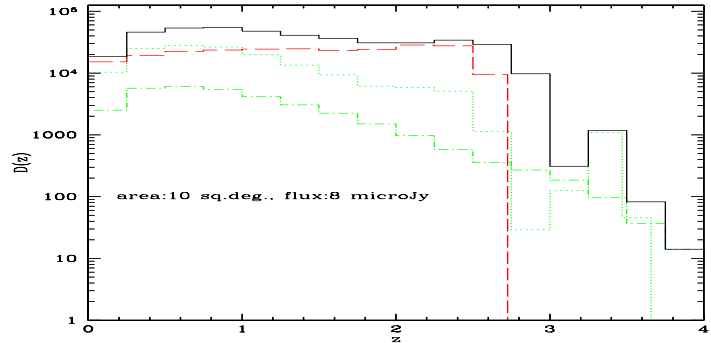}
\end{center}
\caption[]{Predicted redshift distribution for SWIRE-selected galaxies over 10 sq.deg. in the
4.5 $\mu$m channel. Dashed, dotted, and dot-dash lines correspond to E/S0, Sp/Ir and AGN
sub-populations. The evolution model assumes $z_F=2.5$ for E/S0 galaxies.
}
\label{eps3}
\end{figure}

\section{Studying the evolutionary mass function and clustering of galaxies 
with SWIRE}

The two {\sl Legacy Programs} GOODS and SWIRE will provide an overall complementary and
very effective exploitation of SIRTF to study the evolutionary mass function of galaxies
through the analysis of the rest-frame near-infrared spectra.       GOODS will
be able to investigate deep into the galaxy mass function for galaxies up to quite
high redshifts within two pencil-beam volumes centered on HDFN and CDFS.
SWIRE will add to this the spatial dimension, which will be essential to disentangle
the effect of environment on evolution, and to investigate rare classes of sources, like
the most luminous \& massive galaxies at high-z, and populations of strongly clustered
galaxies, like ERO's (Daddi et al. 2001).

Although this capability for spatial sampling will be at the expense of sensitivity,
still SWIRE will have enough of it to detect galaxies down to $M\simeq 10^{10} M_\odot$ 
in the critical redshift interval $1<z<2.5$, over an enormous sky area (67 sq.deg.). 
Fig. \ref{eps2} compares this SWIRE sensitivity with results of current studies 
limited to areas of $\sim 10$ sq.arcmin.
Fig. \ref{eps3} illustrates the potential quality of statistical investigations
based on SWIRE-detected galaxies in any one of the 7 target fields.

\end{document}